\documentclass[pra,twocolumn,showpacs]{revtex4-1}
\usepackage{eurosym}
\usepackage{amsmath,amssymb,amstext}
\usepackage{ulem}
\usepackage[usenames,dvipsnames]{color}
\usepackage{graphicx}
\usepackage{braket}
\usepackage{natbib}
\usepackage{comment}
\usepackage{dcolumn}
\usepackage[english]{babel}
\usepackage{wasysym}
\usepackage{bm}
\usepackage[colorlinks,bookmarks=false,citecolor=blue,linkcolor=red,urlcolor=blue]{hyperref}

\usepackage{appendix}
\usepackage{textcomp,booktabs}
\usepackage[usenames,dvipsnames]{color}
\usepackage{mathrsfs}
\usepackage{float}
\usepackage{xcolor}
\usepackage{colortbl}
\usepackage{amsmath,amssymb,enumerate,epsfig,bbm,calc,color,ifthen,capt-of}
\makeatletter

\def\CT@@do@color{%
	\global\let\CT@do@color\relax
	\@tempdima\wd\z@
	\advance\@tempdima\@tempdimb
	\advance\@tempdima\@tempdimc
	\advance\@tempdimb\tabcolsep
	\advance\@tempdimc\tabcolsep
	\advance\@tempdima2\tabcolsep
	\kern-\@tempdimb
	\leaders\vrule
	\hskip\@tempdima\@plus  1fill
	\kern-\@tempdimc
	\hskip-\wd\z@ \@plus -1fill }
\makeatother
\begin{document}
\title{Real single-loop cyclic three-level configuration of chiral molecules}
\author{Chong Ye$^{1}$}

\author{Quansheng Zhang$^{1}$}

\author{Yong Li$^{1,2}$}\email{liyong@csrc.ac.cn}

\affiliation{$^1$Beijing Computational Science Research Center, Beijing 100193, China}
\affiliation{$^2$Synergetic Innovation Center for Quantum Effects and Applications, Hunan Normal University, Changsha 410081, China}
\date{\today}
\begin{abstract}
  Single-loop cyclic three-level ($\Delta$-type) configuration of chiral molecules was used for enantio-separation in many theoretical works. Considering the effect of molecular rotation, this simple single-loop configuration is generally replaced by a complicated multiple-loop configuration containing multiple degenerate magnetic sub-levels and the ability of the enantio-separation methods is suppressed. For chiral asymmetric top molecules, we propose a scheme to construct a real single-loop $\Delta$-type configuration with no connections to other states by applying three microwave fields with appropriate polarization vectors and frequencies. With our scheme, the previous theoretical proposals for enantio-separation based on single-loop $\Delta$-type configurations can be experimentally realized when the molecular rotation is considered.

\end{abstract}
\maketitle
\section{Introduction}\label{INTR}
Chiral molecule cannot be superimposed on its mirror image through pure translation and/or rotation. The handedness of the two mirror images (enantiomers) is fundamentally important for the enantiomer-selective of their pharmacological effects~\cite{D1,D2,D3,D4}, biological processes~\cite{B1},
the homo-chirality of life~\cite{HCL}, and even fundamental physics representing systems with broken parity states~\cite{P1}.
Despite this, enantio-separation is an important and challenging task
in chemistry and medicine~\cite{PC1,PC2,PC3,PC4,PC5}.

Solely optical enantio-separation methods
have also been investigated theoretically~\cite{OC1,OC2,OC3,OC4,OC5,IS1,IS2,IS3,RS1,RS2}.
One interesting kind of methods~\cite{IS1,IS2,IS3,RS1,RS2} among them
is based on a system with a cyclic three-level ($\Delta$-type) configuration~\cite{3loop,3loop1}. In general, such a system
driven by electric-dipole transitions is forbidden in natural atoms, but can exist in chiral
molecules and other symmetry-broken systems~\cite{SCP,SCP1,SCP2,SCP3,SCP4,SCP5,SCP6}. For chiral molecules, Kr\'{a}l \textit{et al}.~\cite{IS1,IS2} proposed a system with the $\Delta$-type configuration
to realize enantio-separation in the inner-state space via applying three optical (microwave) fields to invoke both one-photon and two-photon processes in the lowest three vibrational levels.
The product of the three Rabi frequencies will change sign with enantiomer. This leads to the chirality-dependency of the system due to the interference of the one-photon and two-photon processes.
After adiabatical (or diabatical) processes, molecules of different chirality which are initially in their respective ground states are transferred to final levels at different energies~\cite{IS1}. With the $\Delta$-type configuration, the inner-state enantio-separation can also be realized by a purely dynamic transfer process via applying optical ultrashort $\pi/2$ and $\pi$ pulses~\cite{IS3}. Based on the $\Delta$-type configuration, one can realize the spatial enantio-separation via a chirality-dependent generalized Stern-Gerlach effect~\cite{RS1,RS2}.

However, a real gas molecule should involve the subspace of rotational states and each
rotational state may have degenerate magnetic sub-levels. Thus, the ideal single-loop $\Delta$-type configuration~\cite{IS1,IS2,IS3,RS1,RS2,Hirota} would be replaced by a multiple-loop $\Delta$-type configuration~\cite{JCP.137.044313}. For the spatial enantio-separation~\cite{RS1,RS2}, this
effect of the molecular rotation will give birth to the relevant reduction of
the chirality-dependent generalized Stern-Gerlach effect~\cite{JCP.137.044313}.
Very recently, some experimental groups have utilized the multiple-loop $\Delta$-type configuration to realize
the inner-state enantio-separation~\cite{PRL.118.123002} as well as enantio-discrimination~\cite{Nature.497.475,PRL.111.023008,
JPCL.6.196,Angew.Chem.10.1002,PCCP.16.11114,ACI,KK,JPCL.7.341,JCP.142.214201} in gas phase samples. It was pointed out~\cite{PRL.118.123002} that one of the reasons limiting the experimental efficiency
is the appearance of multiple loops. With the multiple-loop $\Delta$-type configuration,
it is not possible to achieve perfect enantio-separation as theoretically proposed
in Refs.~\cite{IS1,IS2,IS3}.
Therefore, constructing a real single-loop $\Delta$-type configuration, with
no connections to other states, is strongly demanded for enantio-separation.

For chiral symmetric top molecules, it was theoretically pointed out in
Ref.~\cite{JCP.137.044313} that under the consideration of the molecular rotation
the real single-loop $\Delta$-type
configuration is prohibited due to the selection rules.
However, many kinds of chiral molecules are of asymmetric tops. The
selection rules of asymmetric tops are different from those of symmetric tops.
In this paper, we aim to present a scheme to construct the real single-loop $\Delta$-type configuration for chiral asymmetric top molecules (instead of the symmetric top ones considered in Ref.~\cite{JCP.137.044313}) under the consideration of molecular rotation.
In order to elucidate our scheme, we assume all
the states are in the vibrational ground state and choose the
working states to be the rotational ground state and other
two higher-energy rotational states. 
Three electromagnetic (optical, microwave, or radio frequency) fields are used to invoke three electric-dipole-allowed transitions among them. With the help of the electric-dipole selection rules, we can realize a real single-loop $\Delta$-type configuration of three single states by appropriately choosing the polarization vectors and the frequencies of the three electromagnetic fields. We also demonstrate that the product of the three corresponding Rabi frequencies will change sign with enantiomer, which guarantees the chirality-dependency of the configuration.

\section{Electric-dipole selection rules for rotational transitions of asymmetric top}\label{SRPL}

General chiral molecules such as $1,2-$propanediol, $1,3-$butanediol, carvone, and menthone are of asymmetric tops. For an asymmetric top molecule, the rotational eigenfunctions are $|J,\tau,M\rangle$ with the angular momentum quantum number $J$, the magnetic quantum number $M$, and $\tau$ running from $-J$ to $J$ in unit steps in the order of increasing energy~\cite{AM}. They can be written as a linear combination of prolate symmetric top eigenfunctions $|J,K,M)$~\cite{AM}:
\begin{align}
|J,\tau,M\rangle=\sum_{K=-J}^{J}A^{J}_{K,\tau}|J,K,M).
\end{align}
The coefficients $A^{J}_{K,\tau}$ are given by solving the static
Schr\"{o}dinger equation of the asymmetric top in the basis of prolate symmetric top eigenfunctions~\cite{AM}. The total wavefunction of the molecule can be described as $|\alpha\rangle=|v_\alpha\rangle|J_\alpha,\tau_\alpha,M_\alpha\rangle$ with the vibrational wavefunction $|v_\alpha\rangle$. For clarity, we have used $|...\rangle$ and $|...)$ to distinguish the asymmetric top and symmetric top eigenfunctions.

We consider a linearly $Z$ polarized or a circularly
polarized (in the $X$-$Y$ plane) electromagnetic field in the space-fixed frame
\begin{align}\label{EqE}
\bm{E}^{s}_{\sigma}=\mathrm{Re}
\{\bm{\varepsilon}^{s}_{\sigma}E^{s}_{\sigma} e^{-i(2\pi\nu t+\varphi_\sigma)}\}.
\end{align}
Any electromagnetic field can be written as a linear combination of them.
Here ``$s$'' indicates the space-fixed frame, and $\bm{\varepsilon}^{s}_{\sigma}$ ($\sigma=0,\pm1$) is the polarization vector of the electromagnetic field. $E^{s}_{\sigma}$, $\nu$,
and $\varphi_{\sigma}$ are, respectively, the field amplitude, the frequency, and the initial phase of the electromagnetic field.
Specifically, $\sigma=0$ corresponds to a linearly $Z$ polarized electromagnetic field with the polarization vector $\bm{\varepsilon}^{s}_{0}=\bm{e}^{s}_Z$; $\sigma=1$ ($\sigma=-1$) corresponds to
a circularly polarized electromagnetic field rotating about the $Z$-axis in the right-hand (left-hand)
sense with the polarization vector $\bm{\varepsilon}^{s}_{1}=(\bm{e}^{s}_X+i\bm{e}^{s}_Y)/\sqrt{2}$ [$\bm{\varepsilon}^{s}_{-1}=-(\bm{e}^{s}_X- i\bm{e}^{s}_Y)/\sqrt{2}$]~\cite{Book1}.

Considering only the electric-dipole-allowed transition between an upper level $|\alpha\rangle$ and a lower level $|\beta\rangle$, the Hamiltonian $\bm{\hat{\mu}}\cdot \bm{E}^{s}_{\sigma}$ in the interaction picture is given as
\begin{align}
{H}^{s}_{\sigma}&=\frac{1}{2}\Omega^{\sigma}_{\alpha\beta}
e^{i[2\pi(f_{\alpha\beta}-\nu)t]}|\alpha\rangle\langle\beta|+h.c., \label{Hs}
\end{align}
where $f_{\alpha\beta} \equiv f_{\alpha}-f_\beta$ ($>0$) is the transition frequency with energies of the two states $hf_{\alpha}$ and $hf_{\beta}$ ($h$ is the Planck constant), and the Rabi frequency is
\begin{equation}\label{RB1}
\Omega^{\sigma}_{\alpha\beta}= {E^{s}_{\sigma}}e^{i\varphi_\sigma}\langle\alpha|\bm{\hat{\mu}}\cdot\bm{\varepsilon}^{s}_{\sigma}|\beta\rangle.
\end{equation}
Here $\bm{\hat{\mu}}$ is the electric dipole operator consisting of a sum over all the (nuclear and electronic) charges, weighted by their position vectors measured from a common origin~\cite{AM}.
We have assumed that $f_{\alpha\beta}$ is close to $\nu$. With this,
the counter-rotating terms like ${\Omega}^{\prime\sigma}_{\alpha\beta}e^{i2\pi(f_{\alpha\beta}+v)t}
|\alpha\rangle\langle \beta|$ and the permanent-dipole terms like
$(\Omega^{\sigma}_{\alpha\alpha}e^{i2\pi vt}+\Omega^{\prime\sigma}_{\alpha\alpha}e^{-i2\pi vt})
|\alpha\rangle\langle \alpha|$ have been ignored in Eq.~(\ref{Hs}) since they oscillate rapidly and will affect little the dynamics of the system.
Here we have defined $\Omega^{\sigma}_{jj}= {E^{s}_{\sigma}}e^{i\varphi_\sigma}\langle j|\bm{\hat{\mu}}\cdot\bm{\varepsilon}^{s}_{\sigma}|j\rangle$ and
 ${\Omega}^{\prime\sigma}_{jj^{\prime}}= {E^{s}_{\sigma}}e^{-i\varphi_\sigma}\langle j|
\bm{\hat{\mu}}\cdot(\bm{\varepsilon}^{s}_{\sigma})^{\ast}|j^{\prime}\rangle
$ with $j,j^{\prime}=\alpha,\beta$.

The components of the electric dipole
in the space-fixed frame $\hat{\mu}_{\sigma}^{s}\equiv\bm{\hat{\mu}}\cdot\bm{\varepsilon}^{s}_{\sigma}$ can be obtained by a rotation from the molecular frame~\cite{AM,JCP.137.044313}
\begin{align}\label{MU1}
\hat{\mu}_{\sigma}^{s}=\sum_{\sigma^{\prime}=\pm1,0}[D^{1}_{\sigma\sigma^{\prime}}
(\psi,
\theta,\phi)]^{\ast}\hat{\mu}^{m}_{\sigma^{\prime}}.
\end{align}
The notation ``$m$'' indicates the molecular frame. $D^{1}$ is the rotation matrix in three dimensions. ``$\ast$'' denotes taking conjugate complex. $\psi$, $\theta$, $\phi$ are the Euler angles connecting the molecular frame and the space-fixed frame. $\hat{\mu}^{m}_{\sigma^{\prime}}$ are the components of the electric dipole in the molecular frame with $\hat{\mu}^{m}_{0}=\hat{\mu}^{m}_z$, $\hat{\mu}^{m}_{+}=
({\hat{\mu}}^{m}_x + i{\hat{\mu}}^{m}_y)/\sqrt{2}$, and $\hat{\mu}^{m}_{-}=
-({\hat{\mu}}^{m}_x - i{\hat{\mu}}^{m}_y)/\sqrt{2}$.
Here $x$, $y$, $z$ are the principal axes of the molecule in the molecular frame. We have used $(X,Y,Z)$ and $(x,y,z)$ to distinguish the coordinates in the space-fixed frame and that in the molecular frame. With Eq.~(\ref{MU1}), the Rabi frequency is
\begin{align}\label{OME2}
\Omega^{\sigma}_{\alpha\beta}
=(-1)^{M_{\beta}+\sigma}
{E}^{s}_{\sigma}e^{i\varphi_\sigma}W^{(\sigma)}_{J_\alpha M_\alpha,J_\beta M_\beta}\Gamma_{J_\alpha\tau_\alpha,J_\beta\tau_\beta},
\end{align}
where the reduced matrix element is
\begin{align}\label{OM3}
&\Gamma_{J_\alpha\tau_\alpha,J_\beta\tau_\beta}=\sqrt{(2J_\alpha+1)(2J_\beta+1)}
\sum_{\sigma^{\prime}=\pm1,0}\langle v_\alpha|\hat{\mu}^{m}_{\sigma^{\prime}}|v_{\beta}\rangle\times\nonumber\\
&\sum^{J_\alpha}_{K_\alpha=-J_\alpha}\sum^{J_\beta}_{K_\beta=-J_\beta}
(-1)^{-K_{\beta}+\sigma^{\prime}}(A^{J_\alpha}_{K_\alpha,\tau_\alpha})^{\ast}
A^{J_\beta}_{K_\beta,\tau_\beta}
W^{(\sigma^{\prime})}_{J_\alpha K_\alpha,J_\beta K_\beta}
\end{align}
and $W^{(\sigma^{\prime\prime})}_{J M,J^{\prime} M^{\prime}}=\left(
  \begin{array}{ccc}
    J & 1 & J^{\prime}\\
    M & -\sigma^{\prime\prime} & -M^{\prime} \\
  \end{array}
\right)$ for $\sigma^{\prime\prime}=0,\pm1$ are 3$j$-symbols. We note that the process to achieve Eq.~(\ref{OME2}) is similar to that in Ref.~\cite{JCP.137.044313} except that we consider the case of asymmetric tops instead of the case of symmetric tops in Ref.~\cite{JCP.137.044313}.

Obviously, 3$j$-symbols play a center role in determining electric-dipole selection rules. For the selection rules of $J$, we have $\Delta J=J_{\alpha}-J_{\beta}=0,\pm1$ according to the 3$j$-symbols in both Eq.~(\ref{OME2}) and Eq.~(\ref{OM3}).
The selection rules of $M$ are directly related to the polarization vectors of the electromagnetic field as demonstrated by the 3$j$-symbol in Eq.~(\ref{OME2}), which gives $\Delta M=M_{\alpha}-M_{\beta}=\sigma$.

In principle, using the Wigner-Eckart theorem, one can write the
Rabi frequency in the form of Eq.~(\ref{OME2}) and easily
get the selection rules of $J$ and $M$. However, the reduced matrix element~(\ref{OM3}) is fundamentally important for constructing the real single-loop $\Delta$-type configuration. This will be seen distinctly
when we simplify our discussion to the case of symmetric tops with reducing Eq.~(\ref{OM3})
\begin{align}\label{OM4}
&\Gamma_{J_\alpha K_\alpha,J_\beta K_\beta}=\sqrt{(2J_\alpha+1)(2J_\beta+1)}\times\nonumber\\
&\sum_{\sigma^{\prime}=\pm1,0}(-1)^{-K_{\beta}+\sigma^{\prime}}\langle v_\alpha|\hat{\mu}^{m}_{\sigma^{\prime}}|v_{\beta}\rangle
W^{(\sigma^{\prime})}_{J_\alpha K_\alpha,J_\beta K_\beta}.
\end{align}
The 3$j$-symbol $W^{(\sigma^{\prime})}_{J_\alpha K_\alpha,J_\beta K_\beta}$ here
establishes the selection rules of $K$. This is one of the reasons for
preventing the formation of the single-loop $\Delta$-type configuration for chiral
symmetric top molecules~\cite{JCP.137.044313}.
For the case of asymmetric tops, the sum over $K_{\alpha}$ and $K_{\beta}$ in Eq.~(\ref{OM3}) releases the selection rules of $K$ and thus offers the possibility of forming the closed single-loop $\Delta$-type configuration for chiral asymmetric top molecules.

\section{Real single-loop $\Delta$-type configuration}
Our task is to establish a scheme to form the real chirality-dependent single-loop $\Delta$-type configuration with the help of discussions in Sec.~\ref{SRPL} for chiral asymmetric top molecules. For simplicity,
we assume all the states are in the vibrational ground state $|v_g\rangle$ (in fact the case of different vibrational states will bring the similar results).
With this, we only take the rotational subspace into consideration and shorten $\langle v_g|\hat{\mu}^{m}_{\sigma^{\prime}}|v_g\rangle$ to
$\mu^{m}_{\sigma^{\prime}}$ in further discussions.

\subsection{General formula}
A natural starting state of the single-loop $\Delta$-type configuration
is the rotational ground state
$|J_a,\tau_a\rangle=|0,0\rangle$ which has no magnetic degeneracy.
We apply three electromagnetic fields to resonantly couple, respectively, with three cyclic
transitions $|0,0\rangle\rightarrow|J_b,\tau_b\rangle\rightarrow|J_c,\tau_c\rangle\rightarrow|0,0\rangle$.
They can be written as  $\bm{E}_{1}=\mathrm{Re}\{\sum_{\sigma=0,\pm1}E_{1,\sigma}\bm{\varepsilon}^{s}_{\sigma}e^{-i(2\pi\nu_1 t+\varphi_{1,\sigma})}\}$,
$\bm{E}_{2}=\mathrm{Re}\{\sum_{\sigma=0,\pm1}E_{2,\sigma}\bm{\varepsilon}^{s}_{\sigma}e^{-i(2\pi\nu_2 t+\varphi_{2,\sigma})}\}$, and
$\bm{E}_{3}=\mathrm{Re}\{\sum_{\sigma=0,\pm1}E_{3,\sigma}\bm{\varepsilon}^{s}_{\sigma}e^{-i(2\pi\nu_3 t+\varphi_{3,\sigma})}\}$ with $\nu_{1}=f_{ba}$, $\nu_{2}=f_{cb}$, and $\nu_{3}=f_{ca}$. According to the selection rules of $J$, we have $J_b=J_c=1$. Ignoring all the transitions that are off-resonantly coupled with the three fields, the total Hamiltonian in the interaction picture in the
rotating-wave approximation can be arranged as
\begin{align}\label{Eq9}
H_{total}=(\frac{\Omega_{1}}{2}|b\rangle\langle a|+\frac{\Omega_{2}}{2}|c\rangle\langle b|+\frac{\Omega_{3}}{2}|c\rangle\langle a|+h.c.)
+H^{\prime},
\end{align}
where $|a\rangle=|0,0,0\rangle$. $H^{\prime}$ in Eq.~(\ref{Eq9}) is uncoupled with $|a\rangle$, when we choose
\begin{align}\label{Eqb}
&|b\rangle=\sin\theta_1\cos\phi_1e^{i\varphi_{1,1}}|1,\tau_b,1\rangle
+\sin\theta_1\sin\phi_1e^{i\varphi_{1,0}}\nonumber\\
&\times|1,\tau_b,0\rangle
+\cos\theta_1e^{i\varphi_{1,-1}}|1,\tau_b,-1\rangle,
\end{align}
and
\begin{align}\label{Eqc}
&|c\rangle=\sin\theta_3\cos\phi_3e^{i\varphi_{3,1}}|1,\tau_c,1\rangle
+\sin\theta_3\sin\phi_3e^{i\varphi_{3,0}}\nonumber\\
&\times|1,\tau_c,0\rangle
+\cos\theta_3e^{i\varphi_{3,-1}}|1,\tau_c,-1\rangle.
\end{align}
Here $\sin\theta_\lambda\cos\phi_\lambda=E_{\lambda,1}/E_{\lambda}$, $\sin\theta_\lambda\sin\phi_\lambda=E_{\lambda,0}/E_{\lambda}$, and $\cos\theta_\lambda=E_{\lambda,-1}/E_{\lambda}$ with $E_{\lambda}=\sqrt{E^2_{\lambda,1}+E^2_{\lambda,0}+E^2_{\lambda,-1}}$
($\lambda=1,2,3$).

With these, $H^{\prime}$ denotes all the other resonantly coupling terms between $|J_b,\tau_b\rangle$ and $|J_c,\tau_c\rangle$. Since $J_b=J_c=1$, both $|b\rangle$ and $| c\rangle$ have two degenerate states, labeled respectively as
$|b^{\prime}\rangle$, $|b^{\prime\prime}\rangle$, $|c^{\prime}\rangle$, and $|c^{\prime\prime}\rangle$.
If the conditions
\begin{align}\label{C1}
&\langle c^{\prime}|H_{total}|b\rangle=0,~~
\langle c^{\prime\prime}|H_{total}|b\rangle=0,\nonumber\\
&\langle c|H_{total}|b^{\prime}\rangle=0,~~
\langle c|H_{total}|b^{\prime\prime}\rangle=0,
\end{align}
are satisfied, $H^{\prime}$
is uncoupled to the Hilbert space $\{|a\rangle,|b\rangle,|c\rangle\}$, and the evolution of
a system initially prepared in the Hilbert space
will be governed by the single-loop $\Delta$-type configuration with Hamiltonian
\begin{align}\label{HSL}
H_{sl}=\frac{1}{2}(\Omega_{1}|b\rangle\langle a|+\Omega_{2}|c\rangle\langle b|+\Omega_{3}|c\rangle\langle a|+H.c.).
\end{align}
The Rabi frequencies
$\Omega_{1}=-{\Gamma_{1\tau_b,00}}E_{1}/{\sqrt{3}}$ and
$\Omega_{3}=-{\Gamma_{1\tau_c,00}}E_{3}/{\sqrt{3}}$
are $nonzero$ when $\Gamma_{1\tau_b,00}\ne 0$ and $\Gamma_{1\tau_c,00}\ne 0$.
The Rabi frequency $\Omega_{2}$ is proportional to $\Gamma_{1\tau_c,1\tau_b}$ and the ratio between them
is determined by the polarization vectors of the three fields.
From the results in Sec.~\ref{SRPL}, we know $\Gamma_{1\tau_b,00}$, $\Gamma_{1\tau_c,00}$, and $\Gamma_{1\tau_c,1\tau_b}$ are irrelevant to the polarization vectors of the three fields.
In order to have a closed $\Delta$-type configuration, we will ensure $\Gamma_{1\tau_b,00}$, $\Gamma_{1\tau_c,00}$, and $\Gamma_{1\tau_c,1\tau_b}$ are $nonzero$ first in the following.

\subsection{$(J,\tau)$-level structure and chirality-dependency }\label{JT}
As demonstrated previously, $J_b=J_c=1$.
In the $J=1$ subspace, there are three rotational states
$|J=1,\tau=-1\rangle=|J=1,K=0)$, $|1,0\rangle=[|1,1)-|1,-1)]/{\sqrt{2}}$, and $|1,1\rangle=[|1,1)+|1,-1)]/{\sqrt{2}}$~\cite{AM}.
According to the rotational Hamiltonian $H_{rot}=h(AJ^2_z+BJ^2_x+CJ^2_y)$
(for the asymmetric top molecule with the rotational constants $A>B>C$ )~\cite{AM},
we have the eigenenergies for the $J=0$ and $J=1$ rotational states as $hf_{J=0,\tau=0}=0$, $hf_{1,-1}=h(B+C)$, $hf_{1,0}=h(A+C)$, and $hf_{1,1}=h(A+B)$.

\begin{figure}[h]
  \centering
  \includegraphics[width=1\columnwidth]{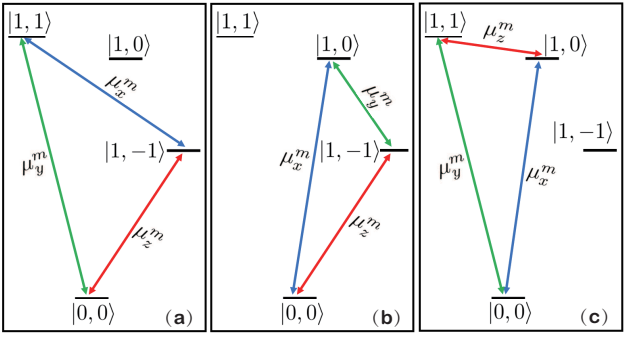}\\
  \caption{ $(J,\tau)$-level structure for the (multiple-loop) $\Delta$-type configuration starting from the rotational ground state with cyclic transitions: (a)   $|J=0,\tau=0\rangle\rightarrow|1,-1\rangle\rightarrow|1,1\rangle\rightarrow|0,0\rangle$;
  (b) $|0,0\rangle\rightarrow|1,-1\rangle\rightarrow|1,0\rangle\rightarrow|0,0\rangle$;
  (c) $|0,0\rangle\rightarrow|1,0\rangle\rightarrow|1,1\rangle\rightarrow|0,0\rangle$. The states are in the $|J,\tau\rangle$ notation. $\mu^{m}_{x}$, $\mu^{m}_y$, and $\mu^{m}_z$ are $nonzero$ electric dipoles in the molecule frame along the principal axis, and respectively proportional to the related reduced matrix elements~(\ref{OM3}).
   }\label{Fig2}
\end{figure}

Three candidates of $(J,\tau)$-level structures for constructing the closed $\Delta$-type configuration are shown in Fig.~\ref{Fig2}. Take the first one [Fig.~\ref{Fig2}(a)] as an example.
It is formed by three cyclic transitions $|0,0\rangle\rightarrow|1,-1\rangle\rightarrow|1,1\rangle\rightarrow|0,0\rangle$ in the $|J,\tau\rangle$ notation.
With Eq.~(\ref{OM3}), we have the $nonzero$ reduced matrix elements~\cite{N1}
\begin{align}\label{RME}
&\Gamma_{1-1,00}=\sqrt{3}\mu^{m}_{0}W^{(0)}_{10,00}
\propto\mu^{m}_{z},\nonumber\\
&\Gamma_{11,1-1}=-\frac{3}{\sqrt{2}}(\mu^{m}_{1}
W^{(1)}_{11,10}+
\mu^{m}_{-1}W^{(-1)}_{1-1,10})\propto\mu^{m}_{x},\nonumber\\
&\Gamma_{11,00}=-\sqrt{\frac{3}{2}}
(\mu^{m}_{1}W^{(1)}_{11,00}
+\mu^{m}_{-1}W^{(-1)}_{1-1,00})\propto\mu^{m}_{y}.
\end{align}
Here $\mu^{m}_x$, $\mu^{m}_y$, and $\mu^{m}_z$ are the $nonzero$ components of the electric dipole in the molecule frame along the respective principal axes. Thus, the $(J,\tau)$-level structure in Fig.~\ref{Fig2}~(a) is available to form the closed $\Delta$-type configuration.

Moreover, the chirality-dependency of the $\Delta$-type configuration is also reflected in the $(J,\tau)$-level structure.
It is known that the sign of $\mu^{m}_x\mu^{m}_y\mu^{m}_z$ fully determines the chirality
of an enantiomer~\cite{Nature.497.475,PRL.111.023008,
JPCL.6.196,PRL.118.123002,Angew.Chem.10.1002,PCCP.16.11114,ACI}.
The sign of any two of the three dipole moment components is arbitrary and changes with the choice of axes, whereas the sign of the combined quantity $\mu^{m}_x\mu^{m}_y\mu^{m}_z$ is axis independent and changes sign with enantiomer.
Combining this with Eq.~(\ref{OME2}), the product of the three reduced matrix elements in Eq.~(\ref{RME})
as well as the product of the three Rabi frequencies in Eq.~(\ref{HSL})
will change sign with enantiomer.
This guarantees the chirality-dependency of the $\Delta$-type configuration.
Applying similar analyses on the other two candidates shown in Fig.~\ref{Fig2}~(b)
and Fig.~\ref{Fig2}~(c), we find that they are also available to form the closed and chirality-dependent $\Delta$-type configuration.

\subsection{$M$-level structure with $Z$ and circularly polarized electromagnetic fields}\label{MSP}

Starting from the rotational ground state, so far we have given three kinds of
$(J,\tau)$-level structures for forming the closed and chirality-dependent $\Delta$-type configuration.
However, due to the magnetic degeneracy of $|J_b=1,\tau_b\rangle$ and $|J_c=1,\tau_c\rangle$, generally such a $\Delta$-type configuration still has multiple loops when the polarizations of the electromagnetic fields are not appropriately chosen. In this subsection, we will turn to the conditions~(\ref{C1}), which provide the selection of
the appropriated polarization vectors of the three electromagnetic fields
to achieve the single-loop $\Delta$-type configuration with the Hamiltonian~({\ref{HSL}}).

We consider the situation where only the linearly $Z$ ($\sigma=0$) or circularly polarized electromagnetic field ($\sigma=\pm 1$) is applied to resonantly couple with each transition in the $\Delta$-type configuration. The three electromagnetic fields are
$\bm{E}_{1}=\mathrm{Re}\{E_{1,\sigma_1}\bm{\varepsilon}^{s}_{\sigma_1}e^{-i(2\pi\nu_1 t+\varphi_{1,\sigma_1})}\}$,
$\bm{E}_{2}=\mathrm{Re}\{E_{2,\sigma_2}\bm{\varepsilon}^{s}_{\sigma_2}e^{-i(2\pi\nu_2 t+\varphi_{2,\sigma_2})}\}$, and
$\bm{E}_{3}=\mathrm{Re}\{E_{3,\sigma_3}\bm{\varepsilon}^{s}_{\sigma_3}e^{-i(2\pi\nu_3 t+\varphi_{3,\sigma_3})}\}$. Here, $\sigma_1$, $\sigma_2$, and $\sigma_3$ stand for their
polarization vectors.

In this case, according to the selection rules of $M$, $\bm{E}_1$, and $\bm{E}_3$ only evoke transitions
$|0,0,0\rangle\rightarrow|1,\tau_b,M_b\rangle$ and $|0,0,0\rangle\rightarrow|1,\tau_c,M_c\rangle$, respectively. Thus, we have $|b\rangle =|J_b=1,\tau_b,M_b\rangle$ and $|c\rangle =|J_c=1,\tau_c,M_c\rangle$. If $\bm{E}_2$
can evoke the transition $|b\rangle\rightarrow|c\rangle$, the conditions~(\ref{C1}) are satisfied according to the selection rules of $M$ and we can
form the real single-loop $\Delta$-type configuration with the Hamiltonian~(\ref{HSL}).

All possible $M$-level structures for constructing the single-loop $\Delta$-type configuration are listed in Table~\ref{Tab1}.
\begin{table}[h]
\begin{tabular}{rrrrrr}
  \hline
   \rowcolor{blue!10}   $~~~~~M_a$ & $~~~~~M_b$ & $~~~~~M_c$ & $~~~~~\sigma_{1}$ & $~~~~~\sigma_{2}$ & $~~~~~\sigma_{3}~~~~~$ \\
  \hline $0$~~ & $1$~~ & 0~~  & $ 1 $~  & $-1$~ & $0$~~~~~~ \\
  \rowcolor{gray!10} 0~~  &$-1$~~ & 0~~  & $-1$~  & $1$~ & $0$~~~~~~ \\
   0~~  &0~~  & 1~~  & $0$~  & $1$~ & $1$~~~~~~~\\
  \rowcolor{gray!10}  0~~  &0~~  & $-1$~~ & $0$~  & $-1$~ &$-1$~~~~~~~\\
   0~~  &$-1$~~  & $-1$~~ & $-1$~  & $0$~ & $-1$~~~~~~~\\
  \rowcolor{gray!10}  0~~  &1~~  & 1~~  & $1$~ & $0$~ & $1$~~~~~~~\\
  \hline
\end{tabular}
  \caption{$M$-level structure to form the real single-loop $\Delta$-type configuration starting from the rotational ground state. $\sigma_{1}$, $\sigma_{2}$, and $\sigma_{3}$ label the polarization vectors of the electromagnetic fields $\bm{E}_1$, $\bm{E}_2$, and $\bm{E}_3$. Here we only choose the $Z$ polarized ($\sigma=0$) electromagnetic field and circularly polarized ($\sigma=\pm$) electromagnetic field rotating about the $Z$-axis. The ($J,\tau$)-level structure of the configuration could be any one of the three cases in Fig.~\ref{Fig2}.
  }\label{Tab1}
\end{table}
We would like to note that, according to the selection rules of $J$ and $M$, it seems the closed single-loop $\Delta$-type configuration among $|0,0,0\rangle$, $|J_{b}=1,\tau_{b},M_b=0\rangle$, and $|J_{c}=1,\tau_{c},M_c=0\rangle$ can be constructed with three linearly $Z$ polarized electromagnetic fields. However, such a configuration will fail since the transition $|J_{b}=1,\tau_{b},M_b=0\rangle \rightarrow|1,\tau_c,0\rangle$ is forbidden by $W^{(0)}_{10,10}=0$.

So far, we can form a closed and chirality-dependent single-loop $\Delta$-type configuration described by the Hamiltonian~(\ref{HSL}) for chiral
asymmetric top molecules. The $(J,\tau)$-level and $M$-level structures for constructing the single-loop $\Delta$-type configuration (as well as
the polarizations of the electromagnetic fields) are given by
Fig.~{\ref{Fig2}} and Table~\ref{Tab1} respectively. In addition, we note that changing the axis of quantization (i.e., change $Z$ to $X$ or $Y$) will give the equivalent configuration to what we form above.

\subsection{$M$-level structure with linearly polarized electromagnetic fields}\label{LPF}
In the recent experiment~\cite{PRL.111.023008} of enantio-separation based on the (multiple-loop) $\Delta$-type configuration, the three
electromagnetic fields are linearly polarized. In this subsection, we consider
this situation of purely linearly polarized fields and will prove the single-loop configuration can be formed only when
the polarization vectors of the three electromagnetic fields are
mutually vertical to each other with the help of the conditions~(\ref{C1}).

Without loss of generality, we can set
$\bm{E}_{1}$ as a linearly $Z$ polarized field. This gives
$|b\rangle=|1,\tau_b,0\rangle$ and
\begin{align}\label{E15}
\sin\theta_{1}\sin\phi_1=\pm 1.
\end{align}
Combining this with the condition $\langle c^{\prime}|H_{total}|b\rangle=0$
and the condition $\langle c|H_{total}|b^{\prime}\rangle=0$, we can
prove that both $\bm{E}_2$ and $\bm{E}_3$ are in the $X-Y$ plane
and vertical to $\bm{E}_1$.

Generally,
we can set $\bm{E}_{2}$ as a linearly $X$ polarized field. This gives
\begin{align}\label{E20}
E_{2,1}e^{i\varphi_{2,1}}=-E_{2,-1}e^{i\varphi_{2,-1}}.
\end{align}
With the condition $\langle c^{\prime\prime}|H_{total}|b\rangle=0$, we
can prove that $\bm{E}_3$ is a linearly $Y$ polarized field and vertical to $\bm{E}_2$.

Changing the definition of the coordinates in the space-fixed frame will
not alter the physical properties. Thus, for a real single-loop
configuration coupled with three linearly polarized fields, we have proven the polarization vectors of the fields must be mutually vertical to each other.

\section{Experimental realization for $1,2$-propanediol}
Now we take $1,2$-propanediol as an example to construct real single-loop $\Delta$-type configurations. The rotational constants and
the components of the electric dipole in the molecule frame for $1,2$-propanediol are $A=8,572.05$\,MHz, $B=3,640.10$\,MHz, $C=2,790.96$\,MHz, $|\mu^{m}_x|=1.916$ Debye, $|\mu^{m}_y|=0.365$ Debye, and $|\mu^{m}_z|=1.201$ Debye with
$1~\mathrm{Debye}=3.33564\times10^{-30} \mathrm{C\cdot m}$ \cite{MST}. The $(J,\tau)$-level structure of the single-loop $\Delta$-type configurations is formed by $|J_a,\tau_a\rangle=|0,0\rangle$, $|J_b,\tau_b\rangle=|1,-1\rangle$, and $|J_c,\tau_c\rangle=|1,1\rangle$ as shown in Fig.~\ref{Fig2} (a).
Three microwave fields are applied to couple resonantly with the transitions among them, respectively, with the corresponding frequencies $\nu_{1}=f_{ba}=6,431.06$\,MHz, $\nu_{2}=f_{cb}=5,781.09$\,MHz, and $\nu_{3}=f_{ca}=12,212.15$\,MHz. In the current related experiment~\cite{PRL.118.123002}, the coupling strengths (about $10$\,MHz) are much less than the detunings (about $1$\,GHz). All the other off-resonantly transitions are largely detuned coupled with these three microwave fields and then can be ignored. Thus, one can form the chirality-dependent $\Delta$-type configurations for $1,2$-propanediol.
\subsection{Single-loop $\Delta$-type configuration with $Z$ and circularly polarized electromagnetic fields}\label{SZC}
\begin{figure}[h]
  \centering
  \includegraphics[width=0.9\columnwidth]{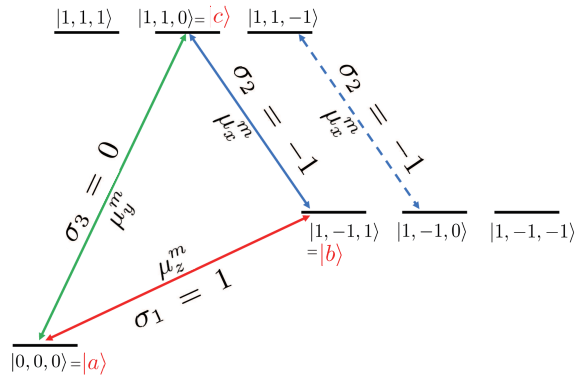}\\
  \caption{Real single-loop $\Delta$-type configuration starting from the rotational ground state for $1,2$-propanediol. Three microwave fields with frequencies $\nu_{1}=6,431.06$\,MHz, $\nu_{2}=5,781.09$\,MHz, $\nu_{3}=12,212.15$\,MHz and
  polarization vectors labeled with $\sigma_{1}=1$, $\sigma_{2}=-1$, $\sigma_{3}=0$ are applied to resonantly couple, respectively, with cyclic electric-dipole-allowed transitions $|a\rangle\rightarrow|b\rangle\rightarrow|c
  \rangle\rightarrow|a\rangle$. The dashed arrow denotes for the transition $|1,-1,0\rangle\rightarrow|1,1,-1\rangle$, which is coupled with the $\sigma_{2}=-1$
  polarized electromagnetic field, and, however, is not involved in the single-loop $\Delta$-type configuration based on $|a\rangle$, $|b\rangle$, and $|c\rangle$.
  }\label{ST}
\end{figure}
In this subsection, we show one of the single-loop $\Delta$-type configurations with choosing the first case of the $M$-level structure in Table~\ref{Tab1}.
The working states are $|a\rangle=|J=0,\tau=0,M=0\rangle$, $|b\rangle=|1,-1,1\rangle$, and $|c\rangle=|1,1,0\rangle$. The polarization vectors of the three microwave fields are chosen according to Table~\ref{Tab1}, labeled with $\sigma_{1}=1$, $\sigma_{2}=-1$, and $\sigma_{3}=0$. Thus
the three microwave fields are $\bm{E_{1}}=\mathrm{Re}
\{\bm{\varepsilon}^{s}_{1}E_{1,1} e^{-i2\pi\nu_1 t}\}$,
$\bm{E_{2}}=\mathrm{Re}
\{\bm{\varepsilon}^{s}_{-1}E_{2,-1} e^{-i2\pi\nu_2 t}\}$,
and $\bm{E_{3}}=\mathrm{Re}
\{\bm{\varepsilon}^{s}_{0}E_{3,0} e^{-i2\pi\nu_3 t}\}$ with
$E_{1,1}>0$, $E_{2,-1}>0$, and $E_{3,0}>0$. For simplicity, we have set the initial phase of them to be $zero$.

For clarity, we show in Fig.~\ref{ST} all the magnetic sub-levels in the $\{|J=0,\tau=0\rangle,|1,-1\rangle,|1,1\rangle\}$ subspace and all the electric-dipole-allowed transitions that are coupled resonantly with the three microwave fields. The rotation states $|1,1,1\rangle$ and $|1,-1,-1\rangle$ are decoupled with the chosen microwave fields. Note that the transition $|1,-1,0\rangle\rightarrow|1,1,-1\rangle$ (the dashed arrow in Fig.~\ref{ST}) is also coupled with the field $\bm{E}_2$. However, it is not involved in the single-loop $\Delta$-type configuration constructed by $|a\rangle=|0,0,0\rangle$, $|b\rangle=|1,-1,1\rangle$, and $|c\rangle=|1,1,0\rangle$ in Fig.~\ref{ST}. The corresponding Rabi frequencies are $\Omega_{1}=-{\Gamma_{1-1,00}}{E}_1/{\sqrt{3}}$, $\Omega_{2}=-{\Gamma_{11,1-1}}{E}_2/{\sqrt{6}}$, and
$\Omega_{3}=-{\Gamma_{11,00}}{E}_3/{\sqrt{3}}$, where
${E}_1=E_{1,1}$, ${E}_2=E_{2,-1}$, and ${E}_3=E_{3,0}$
are the intensities of the three microwave fields.

\subsection{Single-loop $\Delta$-type configuration with linearly polarized electromagnetic fields}
\begin{figure}[h]
  \centering
  \includegraphics[width=0.9\columnwidth]{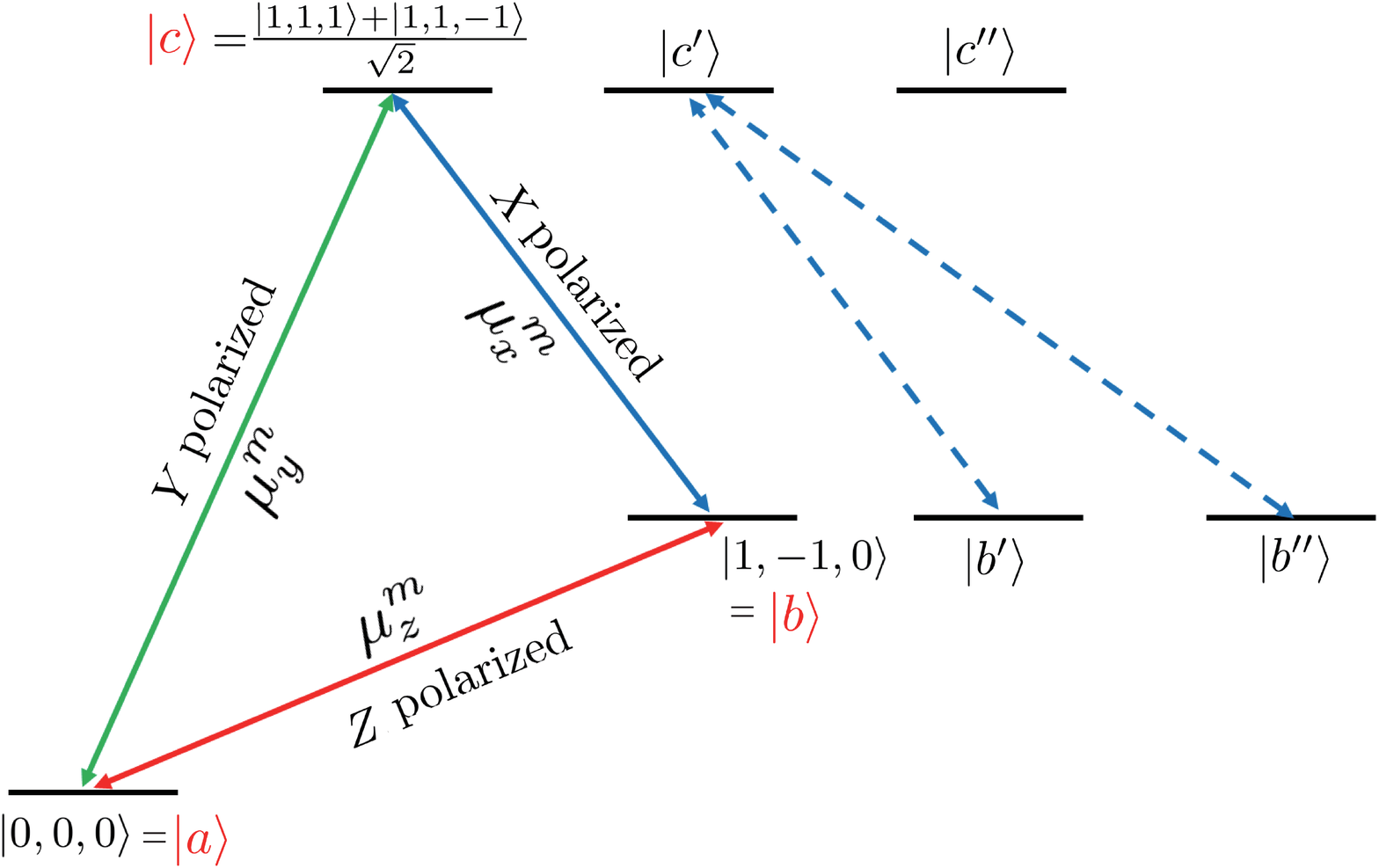}\\
  \caption{Real single-loop $\Delta$-type configuration coupled with three linearly polarized electromagnetic fields for $1,2$-propanediol. Their frequencies are $\nu_{1}=5,781.09$\,MHz, $\nu_{2}=6,431.06$\,MHz, $\nu_{3}=12,212.15$\,MHz.
  Their polarization vectors are $Z$, $X$, and $Y$ respectively. They are resonantly coupled respectively with cyclic electric-dipole-allowed transitions $|a\rangle\rightarrow|b\rangle\rightarrow|c
  \rangle\rightarrow|a\rangle$. Here $|a\rangle=|0,0,0\rangle$, $|b\rangle=|1,-1,0\rangle$, and $|c\rangle=(|1,1,1\rangle+|1,1,-1\rangle)/\sqrt{2}$.
  The dashed arrows denote for the transitions $|b^{\prime}\rangle\rightarrow|c^{\prime}\rangle$ and
  $|b^{\prime\prime}\rangle\rightarrow|c^{\prime}\rangle$, which are coupled with the $X$ polarized field, and, however, are not involved in the single-loop $\Delta$-type configuration based on $|a\rangle$, $|b\rangle$, and $|c\rangle$. Here $|b^{\prime}\rangle=|1,-1,1\rangle$, $|b^{\prime\prime}\rangle=-|1,-1,-1\rangle$, $|c^{\prime}\rangle=-|1,1,0\rangle$, and
  $|c^{\prime\prime}\rangle=(|1,1,1\rangle-|1,1,-1\rangle)/\sqrt{2}$.}\label{Fig3}
\end{figure}
In this subsection, we show an example of the real single-loop configurations resonantly coupled with three linearly polarized electromagnetic fields (as demonstrated in Fig.~\ref{Fig3}) according to the discussions in Sec.~\ref{LPF}.
Here, $\bm{E_{1}}=\mathrm{Re}
\{\bm{\varepsilon}^{s}_{0}E_{1,0} e^{-i2\pi\nu_1 t}\}$ with $E_{1,0}>0$,
$\bm{E_{2}}=\mathrm{Re}
\{(\bm{\varepsilon}^{s}_{1}E_{2,1}+\bm{\varepsilon}^{s}_{-1}E_{2,-1}) e^{-i2\pi\nu_2 t}\}$ with $E_{2,1}=-E_{2,-1}>0$,
and $\bm{E_{3}}=\mathrm{Re}
\{(\bm{\varepsilon}^{s}_{1}E_{3,1}+\bm{\varepsilon}^{s}_{-1}E_{3,-1})e^{-i2\pi\nu_3 t}\}$ with $E_{3,1}=E_{3,-1}>0$
are linearly $Z$, $X$, and $Y$ polarized electromagnetic fields, respectively. For simplicity, we have set the initial phase of them to be $zero$.


With Eq.~(\ref{Eqb}) and Eq.~(\ref{Eqc}), we have the three working states $|a\rangle=|0,0,0\rangle$, $|b\rangle=|1,-1,0\rangle$, and $|c\rangle=(|1,1,1\rangle+|1,1,-1\rangle)/\sqrt{2}$.
The corresponding Rabi frequencies are
$\Omega_{1}=-{\Gamma_{1-1,00}}{E}_1/{\sqrt{3}}$, $\Omega_{2}=-{\Gamma_{11,1-1}}{E}_2/{\sqrt{6}}$,
and $\Omega_{3}=-{\Gamma_{11,00}}{E}_3/{\sqrt{3}}$,
where
${E}_1=E_{1,0}$, ${E}_2=\sqrt{2}E_{2,1}$, and ${E}_3=\sqrt{2}E_{3,1}$
are the intensities of the three microwave fields.
Generally, the three Rabi frequencies should be comparable
to ensure that the dynamics of each transition can affect the global dynamics of
the three-level process. In principle, this can be achieved by adjusting the intensities of the involved microwave fields and/or choosing the appropriate specific three levels. Here the absolute values of the three transition dipole moments are, respectively, given as
 $|\Gamma_{1-1,00}/\sqrt{3}|=|\mu^{m}_{z}|/\sqrt{3}\simeq0.693~\mathrm{Debye}$,
$|\Gamma_{11,1-1}/\sqrt{6}|=|\mu^{m}_{x}|/2\simeq0.958~\mathrm{Debye}$, and
$|\Gamma_{11,00}/\sqrt{3}|=|\mu^{m}_{y}|/\sqrt{3}\simeq0.211~\mathrm{Debye}$.
The three Rabi frequencies can be comparable under current experimental conditions~\cite{PRL.118.123002}, where the ratio of the intensities of the three microwave fields is $1:0.75:2.75$ and correspondingly we can give comparable Rabi frequencies as $|\Omega_1|:|\Omega_2|:|\Omega_3|\simeq 1:1.04:0.84$.
This argument is also suitable for the example in Sec.~\ref{SZC}, where
the Rabi frequencies have the same forms as those of here.

We also give the four states
$|b^{\prime}\rangle=|1,-1,1\rangle$, $|b^{\prime\prime}\rangle=-|1,-1,-1\rangle$, $|c^{\prime}\rangle=-|1,1,0\rangle$, and
$|c^{\prime\prime}\rangle=(|1,1,1\rangle-|1,1,-1\rangle)/\sqrt{2}$. The transitions $|b^{\prime}\rangle\rightarrow|c^{\prime}\rangle$ and
$|b^{\prime\prime}\rangle\rightarrow|c^{\prime}\rangle$ are coupled with the $X$
polarized field. However,
they are not involved in the single-loop $\Delta$-type configuration.

\section{Summary and Discussion}

In conclusion, via appropriately choosing the frequencies and polarization vectors of three applied electromagnetic fields, we have established the scheme to form the closed and chirality-dependent real single-loop $\Delta$-type configuration starting from the rotational ground state for chiral asymmetric top molecules with only electric-dipole-allowed rotational transitions under the consideration of molecular rotation.

With our scheme, we have overcome the impediment to
enantio-separation due to the averaging over the
degenerate magnetic sub-levels. With our
scheme, an inner state will be only occupied
by one of the enantiomers via applying previous theoretical proposals~\cite{IS1,IS2,IS3}. However, there are
other impediments to enantio-separation in practice
such as the temperature and the phase mismatching.
At finite temperate, the system is initially in a thermal equilibrium state.
The population in the upper states ($|b\rangle$ and $|c\rangle$) will execute the cycle ``in reverse''~\cite{JPCL.6.196}.
Extending our results to the cases where different vibrational states
are involved, we can achieve higher population difference between the
states initially driven and thus enantio-separation will be increased.
Since the wave-vectors ($\bm{k}_1$, $\bm{k}_2$, and $\bm{k}_3$)
of the three electromagnetic fields
cannot be parallel, there are inevitable phase mismatching in practice.
It will impede the enantio-separation~\cite{JPCL.6.196}. In our discussion, we
have $|\bm{k}_1|>|\bm{k}_2|>|\bm{k}_3|$. In order to minimize
the effect of the phase mismatching, we should
take $\bm{k}_1$ and $\bm{k}_2$ to be parallel and
$\bm{k}_3$ to be perpendicular to them~\cite{KK}.

%
%
%
In addition, systems with $\Delta$-type configuration
are also used in the enantio-discrimination experiments~\cite{Nature.497.475,PRL.111.023008,
JPCL.6.196,Angew.Chem.10.1002,PCCP.16.11114,ACI,JPCL.7.341,JCP.142.214201}.
The $\Delta$-type configuration used in the experiment~\cite{Nature.497.475} also starts from the rotational ground state and thus is similar to the case we consider here. However, such a configuration in this experiment~\cite{Nature.497.475} is not a single-loop $\Delta$-type one. The upper two levels are off-resonantly coupled by a time-varying electric field. Such an electric field will couple other levels to the configuration. These couplings can
not be ignored. Using our scheme to form a real single-loop $\Delta$-type configuration may help to improve the enantio-discrimination efficiency in experiments.

\section{Acknowledgement}
This work is supported by the National Natural Science Foundation of China (under Grants No. 11774024, No. 11534002, No. U1530401), and the Science Challenge Project (under Grant No. TZ2018003).

\appendix

\section{Calculation of $|b\rangle$ and $|c\rangle$ }
\label{AP1}
For the transition $|J_a,\tau_a\rangle\rightarrow|J_b,\tau_b\rangle$ coupled with $\bm{E}_1$, we have
\begin{align}\label{APH1}
&H_{1}=\frac{1}{2}\Gamma_{1\tau_b,00}(-
{E}_{1,1}e^{i\varphi_{1,1}}W^{(1)}_{11,00}|J_b,\tau_b,1\rangle\langle0,0,0|\nonumber\\
&+{E}_{1,0}e^{i\varphi_{1,0}}W^{(0)}_{1 0,0 0}|J_b,\tau_b,0\rangle\langle0,0,0|-{E}_{1,-1}e^{i\varphi_{1,-1}}W^{(-1)}_{1-1,00}\nonumber\\
&\times|J_b,\tau_b,-1\rangle\langle0,0,0|)+h.c.\nonumber\\
&=-\frac{1}{2}\frac{\Gamma_{1\tau_b,00}}{\sqrt{3}}(
{E}_{1,1}e^{i\varphi_{1,1}}|J_b,\tau_b,1\rangle\langle a|+{E}_{1,0}e^{i\varphi_{1,0}}|J_b,\tau_b,0\rangle\langle a|\nonumber\\
&+{E}_{1,-1}e^{i\varphi_{1,-1}}|J_b,\tau_b,-1\rangle\langle a|)+h.c.\nonumber\\
&=\frac{1}{2}\Omega_{1}|b\rangle\langle a|+h.c.
\end{align}
Here $\Omega_1=-\frac{\Gamma_{1\tau_b,00}}{\sqrt{3}}E_{1}$ and
\begin{align}\label{beq}
&|b\rangle=
\frac{{E}_{1,1}e^{i\varphi_{1,1}}}{E_{1}}|J_b,\tau_b,1\rangle
+\frac{{E}_{1,0}e^{i\varphi_{1,0}}}{E_{1}}|J_b,\tau_b,0\rangle\nonumber\\
&+\frac{{E}_{1,-1}e^{i\varphi_{1,-1}}}{E_{1}}|J_b,\tau_b,-1\rangle.
\end{align}

Making $\sin\theta_1\cos\phi_1=E_{1,1}/E_{1}$, $\sin\theta_1\sin\phi_1=E_{1,0}/E_{1}$,
and $\cos\theta_1=E_{1,-1}/E_{1}$, we have
\begin{align}
&|b\rangle=\sin\theta_1\cos\phi_1e^{i\varphi_{1,1}}|J_b,\tau_b,1\rangle
+\sin\theta_1\sin\phi_1e^{i\varphi_{1,0}}\nonumber\\
&\times|J_b,\tau_b,0\rangle
+\cos\theta_1e^{i\varphi_{1,-1}}|J_b,\tau_b,-1\rangle,
\end{align}
\begin{align}\label{EqB1}
&|b^{\prime}\rangle=\sin\phi_1e^{i\varphi_{1,1}}|J_b,\tau_b,1\rangle
-\cos\phi_1e^{i\varphi_{1,0}}|J_b,\tau_b,0\rangle,
\end{align}
and
\begin{align}\label{EqB2}
&|b^{\prime\prime}\rangle=\cos\theta_1\cos\phi_1e^{i\varphi_{1,1}}|J_b,\tau_b,1\rangle
+\cos\theta_1\sin\phi_1e^{i\varphi_{1,0}}\nonumber\\
&\times|J_b,\tau_b,0\rangle
-\sin\theta_1e^{i\varphi_{1,-1}}|J_b,\tau_b,-1\rangle.
\end{align}

For the transition $|J_a,\tau_a\rangle\rightarrow|J_c,\tau_c\rangle$ coupled with $\bm{E}_3$, we have
\begin{align}
&H_{3}=\frac{1}{2}\Gamma_{1\tau_c,00}(
-{E}_{3,1}e^{i\varphi_{3,1}}W^{(1)}_{1 1,00}|J_c,\tau_c,1\rangle\langle0,0,0|\nonumber\\
&+{E}_{3,0}e^{i\varphi_{3,0}}W^{(0)}_{1 0,00}|J_c,\tau_c,0\rangle\langle0,0,0|\nonumber\\
&-{E}_{3,-1}e^{i\varphi_{3,-1}}W^{(-1)}_{1-1,0 0}|J_c,\tau_c,-1\rangle\langle0,0,0|+h.c.
\end{align}
This can be arranged as
$H_{3}=\frac{1}{2}\Omega_{3}|c\rangle\langle a|+h.c.$
with
$\Omega_3=-\frac{\Gamma_{1\tau_c,00}}{\sqrt{3}}E_{3}$ and
\begin{align}\label{ceq}
&|c\rangle=
\frac{{E}_{3,1}e^{i\varphi_{3,1}}}{E_{3}}|J_c,\tau_c,1\rangle
+\frac{{E}_{3,0}e^{i\varphi_{3,0}}}{E_{3}}|J_c,\tau_c,0\rangle\nonumber\\
&+\frac{{E}_{3,-1}e^{i\varphi_{3,-1}}}{E_{3}}|J_c,\tau_c,-1\rangle.
\end{align}

Making $\sin\theta_3\cos\phi_3=E_{3,1}/E_{3}$, $\sin\theta_3\sin\phi_3=E_{3,0}/E_{3}$,
and $\cos\theta_3=E_{3,-1}/E_{3}$, we have
\begin{align}\label{EqC}
&|c\rangle=\sin\theta_3\cos\phi_3e^{i\varphi_{3,1}}|J_c,\tau_c,1\rangle
+\sin\theta_3\sin\phi_3e^{i\varphi_{3,0}}\nonumber\\
&\times|J_c,\tau_c,0\rangle
+\cos\theta_3e^{i\varphi_{3,-1}}|J_c,\tau_c,-1\rangle,
\end{align}
\begin{align}\label{EqC1}
&|c^{\prime}\rangle=\sin\phi_3e^{i\varphi_{3,1}}|J_c,\tau_c,1\rangle
-\cos\phi_3e^{i\varphi_{3,0}}|J_c,\tau_c,0\rangle,
\end{align}
and
\begin{align}\label{EqC2}
&|c^{\prime\prime}\rangle=\cos\theta_3\cos\phi_3e^{i\varphi_{3,1}}|J_c,\tau_c,1\rangle
+\cos\theta_3\sin\phi_3e^{i\varphi_{3,0}}\nonumber\\
&\times|J_c,\tau_c,0\rangle
-\sin\theta_3e^{i\varphi_{3,-1}}|J_c,\tau_c,-1\rangle.
\end{align}

\section{Specific expression of $\langle c|H_{total}|b^{\prime}\rangle$ etc.}\label{AP2}
In order to calculate the specific expression of $\langle c|H_{total}|b^{\prime}\rangle$ etc.,
we first give the Hamiltonian for the transition $|J_b,\tau_b\rangle\rightarrow|J_c,\tau_c\rangle$ coupled with $\bm{E}_2$.
It can be expressed as
\begin{align}
H_{2}=\sum_{\sigma=0,\pm1}H_{2,\sigma}.
\end{align}
Here
\begin{align}
&H_{2,1}=\frac{1}{2}E_{2,1}e^{i\varphi_{2,1}}\Gamma_{1\tau_c,1\tau_b}(W^{(1)}_{1 0,1 -1}|J_c,\tau_c,0\rangle\langle J_b,\tau_b,-1|\nonumber\\
&-W^{(1)}_{1 1,1 0}|J_c,\tau_c,1\rangle\langle J_b,\tau_b,0|)+h.c.\nonumber\\
&=-\frac{E_{2,1}e^{i\varphi_{2,1}}\Gamma_{1\tau_c,1\tau_b}}{2\sqrt{6}}(|J_c,\tau_c,0\rangle\langle J_b,\tau_b,-1|\nonumber\\
&+|J_c,\tau_c,1\rangle\langle J_b,\tau_b,0|)+h.c.,
\end{align}
\begin{align}
&H_{2,0}=-\frac{1}{2}E_{2,0}e^{i\varphi_{2,0}}\Gamma_{1\tau_c,1\tau_b}(W^{(0)}_{1 -1,1 -1}|J_c,\tau_c,-1\rangle\langle J_c,\tau_b,-1|\nonumber\\
&+W^{(0)}_{1 1,1 1}|J_c,\tau_c,1\rangle\langle J_c,\tau_b,1|)+h.c.\nonumber\\
&=-\frac{E_{2,0}e^{i\varphi_{2,0}}\Gamma_{1\tau_c,1\tau_b}}{2\sqrt{6}}(|J_c,\tau_c,-1\rangle\langle J_c,\tau_b,-1|\nonumber\\
&-|J_c,\tau_c,1\rangle\langle J_c,\tau_b,1|)+h.c.,
\end{align}
and
\begin{align}
&H_{2,-1}=\frac{1}{2}{E}_{2,-1}e^{i\varphi_{2,-1}}\Gamma_{1\tau_c,1\tau_b}(-W^{(-1)}_{1 -1,1 0}|J_c,\tau_c,-1\rangle\langle J_c,\tau_b,0|\nonumber\\
&+W^{(-1)}_{1 0,1 1}|J_c,\tau_c,0\rangle\langle J_c,\tau_b,1|)+h.c.\nonumber\\
&=\frac{{E}_{2,-1}e^{i\varphi_{2,-1}}\Gamma_{1\tau_c,1\tau_b}}{2\sqrt{6}}(|J_c,\tau_c,-1\rangle\langle J_c,\tau_b,0|\nonumber\\
&+|J_c,\tau_c,0\rangle\langle J_c,\tau_b,1|)+h.c..
\end{align}
Thus, we have
\begin{align}\label{bceq}
&H_{2}=\frac{\Gamma_{1\tau_c,1\tau_b}E_{2}}{2\sqrt{6}}(-\sin\theta_2\cos\phi_2e^{i\varphi_{2,1}}
|J_c,\tau_c,0\rangle\langle J_b,\tau_b,-1|\nonumber\\
&-\sin\theta_2\cos\phi_2e^{i\varphi_{2,1}}|J_c,\tau_c,1\rangle\langle J_b,\tau_b,0|\nonumber\\
&-\sin\theta_2\sin\phi_2e^{i\varphi_{2,0}}|J_c,\tau_c,-1\rangle\langle J_c,\tau_b,-1|\nonumber\\
&+\sin\theta_2\sin\phi_2e^{i\varphi_{2,0}}|J_c,\tau_c,1\rangle\langle J_c,\tau_b,1|\nonumber\\
&+\cos\theta_2e^{i\varphi_{2,-1}}|J_c,\tau_c,-1\rangle\langle J_c,\tau_b,0|\nonumber\\
&+\cos\theta_2e^{i\varphi_{2,-1}}|J_c,\tau_c,0\rangle\langle J_c,\tau_b,1|)+h.c..
\end{align}
Here, we use $\sin\theta_2\cos\phi_2=E_{2,1}/E_{2}$, $\sin\theta_2\sin\phi_2=E_{2,0}/E_{2}$,
and $\cos\theta_2=E_{2,-1}/E_{2}$.
We have
\begin{align}\label{F1}
&\langle c|H_{total}|b\rangle=\langle c|H_{2}|b\rangle=\frac{\Gamma_{1\tau_c,1\tau_b}E_{2}}{2\sqrt{6}}\times\nonumber\\
&[-\cos\theta_1\sin\theta_2\cos\phi_2\sin\theta_3\sin\phi_3e^{i(\varphi_{1,-1}+\varphi_{2,1}-\varphi_{3,0})}\nonumber\\
&-\cos\theta_1\sin\theta_2\sin\phi_2\cos\theta_3e^{i(\varphi_{1,-1}+\varphi_{2,0}-\varphi_{3,-1})}\nonumber\\
&-\sin\theta_1\sin\phi_1\sin\theta_2\cos\phi_2\sin\theta_3\cos\phi_3e^{i(\varphi_{1,0}+\varphi_{2,1}-\varphi_{3,1})}\nonumber\\
&+\sin\theta_1\sin\phi_1\cos\theta_2\cos\theta_3e^{i(\varphi_{1,0}+\varphi_{2,-1}-\varphi_{3,-1})}\nonumber\\
&+\sin\theta_1\cos\phi_1\sin\theta_2\sin\phi_2\sin\theta_3\cos\phi_3e^{i(\varphi_{1,1}+\varphi_{2,0}-\varphi_{3,1})}\nonumber\\
&-\sin\theta_1\cos\phi_1\cos\theta_2\sin\theta_3\sin\phi_3e^{i(\varphi_{1,1}+\varphi_{2,-1}-\varphi_{3,0})}]\nonumber\\
&=\frac{1}{2}\Omega_2,
\end{align}
\begin{align}\label{F2}
&\langle c^{\prime}|H_{total}|b\rangle=\langle c^{\prime}|H_{2}|b\rangle=\frac{\Gamma_{1\tau_c,1\tau_b}E_{2}}{2\sqrt{6}}\times\nonumber\\
&[\cos\theta_1\sin\theta_2\cos\phi_2\cos\phi_3e^{i(\varphi_{1,-1}+\varphi_{2,1}-\varphi_{3,0})}\nonumber\\
&-\sin\theta_1\sin\phi_1\sin\theta_2\cos\phi_2\sin\phi_3e^{i(\varphi_{1,0}+\varphi_{2,1}-\varphi_{3,1})}\nonumber\\
&+\sin\theta_1\cos\phi_1\sin\theta_2\sin\phi_2\sin\phi_3e^{i(\varphi_{1,1}+\varphi_{2,0}-\varphi_{3,1})}\nonumber\\
&-\sin\theta_1\cos\phi_1\cos\theta_2\cos\phi_3e^{i(\varphi_{1,1}+\varphi_{2,-1}-\varphi_{3,0})}]
\nonumber\\
&=0,
\end{align}
\begin{align}\label{F3}
&\langle c^{\prime\prime}|H_{total}|b\rangle=\langle c^{\prime\prime}|H_{2}|b\rangle=\frac{\Gamma_{1\tau_c,1\tau_b}E_{2}}{2\sqrt{6}}\times\nonumber\\
&[-\cos\theta_1\sin\theta_2\cos\phi_2\cos\theta_3\sin\phi_3e^{i(\varphi_{1,-1}+\varphi_{2,1}-\varphi_{3,0})}\nonumber\\
&+\cos\theta_1\sin\theta_2\sin\phi_2\sin\theta_3e^{i(\varphi_{1,-1}+\varphi_{2,0}-\varphi_{3,-1})}\nonumber\\
&-\sin\theta_1\sin\phi_1\sin\theta_2\cos\phi_2\cos\theta_3\cos\phi_3e^{i(\varphi_{1,0}+\varphi_{2,1}-\varphi_{3,1})}\nonumber\\
&-\sin\theta_1\sin\phi_1\cos\theta_2\sin\theta_3e^{i(\varphi_{1,0}+\varphi_{2,-1}-\varphi_{3,-1})}\nonumber\\
&+\sin\theta_1\cos\phi_1\sin\theta_2\sin\phi_2\cos\theta_3\cos\phi_3e^{i(\varphi_{1,1}+\varphi_{2,0}-\varphi_{3,1})}
\nonumber\\
&+\sin\theta_1\cos\phi_1\cos\theta_2\cos\theta_3\sin\phi_3e^{i(\varphi_{1,1}+\varphi_{2,-1}-\varphi_{3,0})}]
\nonumber\\
&=0,
\end{align}
\begin{align}\label{F4}
&\langle c|H_{total}|b^{\prime}\rangle=\langle c|H_{2}|b^{\prime}\rangle=\frac{\Gamma_{1\tau_c,1\tau_b}E_{2}}{2\sqrt{6}}\times\nonumber\\
&[\cos\phi_1\sin\theta_2\cos\phi_2\sin\theta_3\cos\phi_3e^{i(\varphi_{1,0}+\varphi_{2,1}-\varphi_{3,1})}\nonumber\\
&-\cos\phi_1\cos\theta_2\cos\theta_3e^{i(\varphi_{1,0}+\varphi_{2,-1}-\varphi_{3,-1})}\nonumber\\
&+\sin\phi_1\sin\theta_2\sin\phi_2\sin\theta_3\cos\phi_3e^{i(\varphi_{1,1}+\varphi_{2,0}-\varphi_{3,1})}\nonumber\\
&-\sin\phi_1\cos\theta_2\sin\theta_3\sin\phi_3e^{i(\varphi_{1,1}+\varphi_{2,-1}-\varphi_{3,0})}]
\nonumber\\
&=0,
\end{align}
\begin{align}\label{F5}
&\langle c|H_{total}|b^{\prime\prime}\rangle=\langle c|H_{2}|b^{\prime\prime}\rangle=\frac{\Gamma_{1\tau_c,1\tau_b}E_{2}}{\sqrt{6}}\times\nonumber\\
&[\sin\theta_1\sin\theta_2\cos\phi_2\sin\theta_3\sin\phi_3e^{i(\varphi_{1,-1}+\varphi_{2,1}-\varphi_{3,0})}\nonumber\\
&+\sin\theta_1\sin\theta_2\sin\phi_2\cos\theta_3e^{i(\varphi_{1,-1}+\varphi_{2,0}-\varphi_{3,-1})}\nonumber\\
&-\cos\theta_1\sin\phi_1\sin\theta_2\cos\phi_2\sin\theta_3\cos\phi_3e^{i(\varphi_{1,0}+\varphi_{2,1}-\varphi_{3,1})}\nonumber\\
&+\cos\theta_1\sin\phi_1\cos\theta_2\cos\theta_3e^{i(\varphi_{1,0}+\varphi_{2,-1}-\varphi_{3,-1})}\nonumber\\
&+\cos\theta_1\cos\phi_1\sin\theta_2\sin\phi_2\sin\theta_3\cos\phi_3e^{i(\varphi_{1,1}+\varphi_{2,0}-\varphi_{3,1})}\nonumber\\
&-\cos\theta_1\cos\phi_1\cos\theta_2\sin\theta_3\sin\phi_3e^{i(\varphi_{1,1}+\varphi_{2,-1}-\varphi_{3,0})}]
\nonumber\\
&=0.
\end{align}

\section{$M$-level structure with linearly polarized
electromagnetic fields}\label{AP3}

With $\langle c^{\prime}|H_{total}|b\rangle=0$,
we have
\begin{align}\label{E16}
\sin\theta_2\cos\phi_2\sin\phi_3=0.
\end{align}
This gives $E_{2,1}\propto\sin\theta_2\cos\phi_2=0$ or $E_{3,0}\propto\sin\phi_3=0$.
If $E_{2,1}=0$, $E_{2,-1}$ should be equal to $zero$. Otherwise,
$\bm{E}_2$ will not be a linearly polarized field.
In this case, we have $|c\rangle=|1,\tau_c,0\rangle$. However, the transition $|1,\tau_b,0\rangle\rightarrow
|1,\tau_c,0\rangle$ is prohibited. Therefore, we have
to choose \begin{align}\label{E17}
E_{3,0}\propto\sin\phi_3=0,~{E}_{2,1}\propto\sin\theta_2\cos\phi_2\ne 0.
\end{align}
That means $\bm{E}_3$ should be in the $X-Y$ plane.

With the above results and the condition $\langle c|H_{total}|b^{\prime}\rangle=0$
in Eq. (\ref{F4}), we have
\begin{align}\label{E18}
\sin\theta_2\sin\phi_2\sin\theta_3=0.
\end{align}
If $\sin\theta_{3}=0$,
we have $E_{3,1}=0$.  This conflicts with the fact that $E_{3,0}=0$
[see Eq. (\ref{E16})] and $E_3$ is not a linearly polarized field. Thus we have to
make
\begin{align}\label{E19}
E_{2,0}\propto\sin\theta_2\sin\phi_2=0,~E_{3,1}\propto\sin\theta_{3}\ne0.
\end{align}
That means $\bm{E}_2$ is also in the $X-Y$ plane.
With $\sin\phi_3=0$ and $\sin\theta_2\sin\phi_2=0$, the condition $\langle c|H_{total}|b^{\prime\prime}\rangle=0$ in
Eq. (\ref{F5}) is satisfied.

Now, we have proven $\bm{E}_2$ and $\bm{E}_3$ are in the $X-Y$
plane. Thus they are vertical to $\bm{E}_1$. Generally,
we can set $\bm{E}_{2}$ as a linearly $X$ polarized field. This gives
\begin{align}\label{E20}
E_{2,1}e^{i\varphi_{2,1}}=-E_{2,-1}e^{i\varphi_{2,-1}}.
\end{align}
With the above results and $\langle c^{\prime\prime}|H_{total}|b\rangle=0$, we have
\begin{align}\label{F33}
E_{3,-1}e^{i\varphi_{3,-1}}=E_{3,1}e^{i\varphi_{3,1}}.
\end{align}
Therefore, $\bm{E}_3$ is a linearly $Y$ polarized field and vertical to $\bm{E}_{2}$.

\end{document}